\documentclass[amsmath,amssymb]{revtex4}

\begin{document}

\title{The local energy production rates of GRB photons and of UHECRs}
\author{Eli Waxman}
\affiliation{Particle Phys. \& Astorphys. dept., Weizmann Institute, Rehovot 76100, Israel}

\begin{abstract}
In a recent analysis \cite{Eichler_Guetta10} it was found that the local ($z=0$) rate at which gamma-ray bursts (GRBs) produce energy in $\sim1$~MeV photons, $Q_{\rm MeV, GRB}^{z=0}$, is $\sim10^{2.5}$ times lower than the local energy production rate in ultra-high energy cosmic-rays (UHECRs). This may appear to be in contradiction with earlier results, according to which $Q_{\rm MeV, GRB}^{z=0}$ is similar to the local energy production rate in $>10^{19}$~eV CRs, $Q_{10\rm EeV}^{z=0}$. This short 
note 
identifies the origin of the apparent discrepancy and shows that $Q_{\rm MeV, GRB}^{z=0}\approx Q_{10\rm EeV}^{z=0}$ holds.
\end{abstract}

\maketitle

Assuming that the highest energy cosmic-rays are protons of extra-Galactic (XG) origin, UHECR measurements imply a local energy production rate, per logarithmic proton energy interval, of $\varepsilon^2d\dot{n}_{p,\rm XG}/d\varepsilon=(5\pm2)\times10^{43}{\rm erg/ Mpc^{3}yr}$ at $\varepsilon>10^{19.5}$~eV \cite{Katz_UHECR_09}. A simple model with $\varepsilon^2d\dot{n}_{p,\rm XG}/d\varepsilon= Q_{10\rm EeV}^{z=0}=5\times10^{43}{\rm erg/ Mpc^{3}yr}$ and a transition from Galactic to XG sources at $10^{19}$~eV is consistent with observations (In such a model, the Galactic flux is comparable to the XG one at $10^{19}$~eV, and negligible at $>10^{19.5}$~eV) \cite{Katz_UHECR_09}. $Q_{10\rm EeV}^{z=0}$ is similar to $Q_{\rm MeV, GRB}^{z=0}$ (assuming $Q_{\rm GRB}$ follows the SFR, i.e $Q_{\rm GRB}^{z=0}\ll Q_{\rm GRB}^{z=1.5}$)
\cite{W04_GRB_CR}.
In \cite{Eichler_Guetta10} it is found that the UHECR production rate exceeds $Q_{\rm MeV, GRB}^{z=0}$ by a factor of $\sim10^{2.5}$ due to a combination of 3 factors.\\
$[i]$ The inferred XG proton production rate increases if one assumes that the Galactic-XG transition occurs at energy $\varepsilon\ll10^{19}$~eV. The inferred production rate under this assumption is $\varepsilon^2d\dot{n}/d\varepsilon\approx 1.7\times10^{45}(\varepsilon/10^{18}{\rm eV})^{-0.7}{\rm erg/ Mpc^{3}yr}$ \cite{Berezinsky08}. In \cite{Eichler_Guetta10} it is assumed that the XG component dominates at $4\times10^{18}$~eV, inferring a production rate of $10^{44.5}{\rm erg/ Mpc^{3}yr}$ ($1/2$ the value inferred in \cite{Berezinsky08}).\\
$[ii]$ The XG CR production rate, $Q_{\rm XG}^{z=0}$, is assumed in \cite{Eichler_Guetta10} to be $\sim10$ times larger than the UHECR production rate, arguing that the generation spectrum of the XG protons extends to energy $\ll10^{18}$~eV. This brings the rate inferred in \cite{Eichler_Guetta10} to $Q_{\rm XG}^{z=0}\approx10^{45.5}{\rm erg/ Mpc^{3}yr}\approx10^2 Q_{10\rm EeV}^{z=0}\approx10^2 Q_{\rm MeV, GRB}^{z=0}$.\\
$[iii]$ $Q_{\rm MeV, GRB}^{z=0}=5\times10^{42}{\rm erg/ Mpc^{3}yr}$ inferred in \cite{Eichler_Guetta10} is $\sim10$ times lower than earlier estimates (e.g. \cite{W04_GRB_CR,Le07}). This may be partly due to an underestimate (by a factor of $\approx2$) of the measured time averaged GRB photon flux (using measurements of FERMI's GBM which is less sensitive than BATSE \cite{BATSE4}), and to not taking into account that a significant part of the GRB photon flux is "missed" due to non-triggered distant/faint bursts. Correcting $Q_{\rm MeV, GRB}^{z=0}$ for this effect depends on the model assumed for the GRB luminosity function and its evolution. It is expected to be of order of a few since at redshift $z=1.5$, beyond which half the detected bursts lie, the threshold flux (required to trigger BATSE) corresponds to a luminosity of $L_{50-300\rm keV}=10^{51}{\rm erg/s}$, similar to the average GRB peak luminosity.

Correcting the factors described in [iii], the analysis of \cite{Eichler_Guetta10} would yield $Q_{\rm MeV, GRB}^{z=0}\approx Q_{10\rm EeV}^{z=0}$. The authors of \cite{Eichler_Guetta10} find $Q_{\rm XG}\gg Q_{10\rm EeV}$, and hence also $Q_{\rm XG}\gg Q_{\rm MeV, GRB}$, mainly 
due to assuming that (i) the Galactic-XG transition occurs well below $10^{19}$~eV and (ii) the XG generation spectrum extends over many decades below $10^{18}$~eV. The former assumption is motivated mainly by the argument that it allows one to explain the $4\times10^{18}$~eV spectral feature by pair production (in proton interactions with the CMB) \cite{Berezinsky08}. The transition energy in such models is well below $4\times10^{18}$~eV, which indeed implies $Q_{\rm XG}\gg Q_{\rm MeV, GRB}$. As explained in \cite{Katz_UHECR_09}, a Galactic-XG transition at $\sim10^{18}$~eV requires fine tuning (of the Galactic and XG contributions)
and is disfavored by the data: It requires that Auger systematically underestimates the energy of the events by 40\% (well above the stated uncertainty) and it requires $d\dot{n}_{p,\rm XG}/d\varepsilon\propto \varepsilon^{-2.7}$, which is inconsistent with the $>10^{19}$~eV data.

Assuming a Galactic-XG transition at $10^{19}$~eV, a "bolometric correction" may still be required in order to estimate $Q_{\rm XG}^{z=0}$ if the XG generation spectrum extends over many decades below $10^{19}$~eV, $Q_{\rm XG}^{z=0}/Q_{10\rm EeV}^{z=0}\sim10$. Note, however, that estimating the ratio of $Q_{\rm XG}^{z=0}$ to the total photon energy production by GRBs, $Q_{\gamma,\rm GRB}^{z=0}$, as
$Q_{\rm XG}^{z=0}/Q_{\gamma,\rm GRB}^{z=0}=Q_{\rm XG}^{z=0}/Q_{\rm MeV, GRB}^{z=0}\sim10$ is quite uncertain. This is due to uncertainties in the redshift evolution of the GRB rate and luminosity function, in the "bolometric correction" for the CR production rate, and in the bolometric correction, $Q_{\gamma,\rm GRB}^{z=0}/Q_{\rm MeV, GRB}^{z=0}>1$, that should also be applied to the photons.

\end{document}